# Variance Estimation in Ranked Set Sampling Using a Concomitant Variable

Ehsan Zamanzade[a,1], and Michael Vock[b]

[a] Department of Statistics, University of Isfahan, Isfahan 81746-73441, Iran.
[b] Institute of Mathematical Statistics and Actuarial Science, University of Bern, Sidlerstrasse 5, CH-3012 Bern, Switzerland.

**Abstract:** We propose a nonparametric variance estimator when ranked set sampling (RSS) and judgment post stratification (JPS) are applied by measuring a concomitant variable. Our proposed estimator is obtained by conditioning on observed concomitant values and using nonparametric kernel regression.

## 1. Introduction

Ranked set sampling (RSS), proposed by McIntyre (1952, 2005), is a sampling strategy which uses ranking information to give more efficient statistical inference than simple random sampling (SRS). To collect a balanced ranked set sample using set size $k$, one first draws a sample of size $k^2$ and then divides it into $k$ samples of size $k$ and ranks them in an increasing magnitude (without actually measuring them, i.e., by eye inspection or using a concomitant variable). One then selects for measurement the observation with rank $r$ from the $r^{th}$ sample, for $r = 1, …, k$. This process is repeated $n$ times in order to obtain a sample of measured units of size $N=nk$. Therefore, a balanced ranked set sample consists of $n$ independent measured units with judgment rank one, $n$ independent measured units with judgment rank two, and so on. An unbalanced ranked set sample differs from the balanced one by the number of measured units with rank $r$ not necessarily being the same for all ranks. In fact, if $n_r$ is the number of observations with rank $r$, then $N = \sum_{r=1}^{k} n_r$ is the total sample size.

Another variation of RSS, proposed by MacEachern et al. (2004), is judgment post stratification (JPS). To collect a JPS sample of size $N$, using set size $k$, one first draws a simple random sample of size $N$ and measures all $N$ units. Then, for each measured unit, one draws a supplemental random sample of size ($k$-1) from the population and finds the rank of the measured unit when it is added to this sample. Therefore, a JPS sample of size $N$ consists of a simple random sample of size $N$ and their corresponding ranks. The similarity of JPS sampling scheme and unbalanced RSS is that the number of units with rank $r$ ($n_r$) is not constant. However, JPS differs from unbalanced RSS by the fact that the $n_r$`s are not fixed in advance. In fact, if $(n_1, n_2, \cdots, n_k)$ is the vector of the numbers of units with rank $i$, then one can simply show that $(n_1, n_2, \cdots, n_k) \sim \text{Multinomial}\left(N, \left(\frac{1}{k}, \frac{1}{k}, \cdots, \frac{1}{k}\right)\right)$.

---

[1] Corresponding author.
 Email addresses: Ehsanzamanzadeh@yahoo.com; E.Zamanzade@sci.ui.ac.ir (E. Zamanzade),
Michael.vock@stat.unibe.ch (M. Vock)

A lot of research has been done on RSS and JPS since their introductions. In the RSS scheme, Takahasi and Wakimoto (1968) were the first who proved that the mean estimator from RSS is more efficient than that from SRS. Stokes (1980), MacEachern et al. (2002), Perron and Sinha (2004) proposed different variance estimators. The problem of estimating a distribution function has been considered by Stokes and Sager (1988), Kvam and Samaniego (1994), Duembgen and Zamanzade (2013). Frey et al. (2007) and Li and Balakrishnan (2008) proposed some tests for assessing the assumption of prefect rankings, followed by Vock and Balakrishnan (2011), Zamanzade et al. (2012), Frey and Wang (2013), and Zamanzade et al. (2014).

In the JPS sampling scheme, Wang et al. (2008), Frey and Feeman (2012) proposed some mean estimators. The problem of estimating the population variance has been considered by Frey and Feeman (2013). Frey and Oztürk (2011), Wang et al. (2012) proposed some distribution function estimators.

Frey (2011) proposed some mean estimators in RSS and JPS based on measuring a concomitant variable, showing how the values of the concomitant variable can be used more efficiently than just for ranking. A good review on existing literature on RSS and its variations is given by Wolfe (2012).

The rest of this paper is organized as follows: In Section 2, we propose a nonparametric variance estimator for ranked set samples and judgment post stratification based on a concomitant variable. Then, in Section 3, we compare the proposed estimator with its leading competitors in the literature. Some concluding remarks are provided in Section 4.

## 2. Introduction of the variance estimator

In this section, we introduce a new variance estimator for ranked set samples and judgment post stratified data in the case that the ranking is based on the measurement of a concomitant variable. Let $Y, X$ be the variable of interest and the concomitant variable, respectively, let $\mathbf{X} = \{X_1, X_2, \cdots, X_{kN}\}$ be the full set of $X$ values which is used in rankings, and let $X_{(1)}^m < X_{(2)}^m < \cdots < X_{(N)}^m$ be the $X$ values corresponding to the measured units ($X_{(i)}^m$ to $Y_i$).

The concomitant variable information can be incorporated into the estimation of the population variance by using the identity $Var(Y) = E(\mu_X^2) - E^2(\mu_X^1)$, where $\mu_X^l = E(Y^l \mid X)$, for $l = 1, 2$. By the identities $E(Y^l) = E(\mu_X^l)$ and $Var(Y^l) = E(Var(Y^l \mid X)) + Var(\mu_X^l)$, for $l = 1, 2$, the estimates of $E(\mu_X^l)$ ($l = 1, 2$) can be used for the estimation of $E(Y^l)$ ($l = 1, 2$), and the estimates based on this conditioning are expected to have less variance than direct estimates.

The parameters $E(\mu_X^l)$ ($l = 1, 2$) can be estimated by taking the average over the $kN$ estimates of $E(Y^l \mid X_{(i)})$, $i = 1, 2, \cdots, kN$, $l = 1, 2$. We propose to estimate the quantities $E(Y^l \mid X = x)$ ($l = 1, 2$) by using nonparametric kernel regression. Let $k(.)$ be a kernel function and $h > 0$ the bandwidth, then these quantities can be estimated using the weighted average:

$$m(x, h_l) = \frac{\sum_{i=1}^{N} Y_i^l k\left(\frac{x - X_{(i)}^m}{h_l}\right)}{\sum_{i=1}^{N} k\left(\frac{x - X_{(i)}^m}{h_l}\right)}, \quad l = 1, 2.$$

We use the standard normal density function as the kernel function, and since we want to use the regression equation for prediction, we propose to select the value for the bandwidth based on the cross-validation (CV) method. I. e., we select the value of the bandwidth ($h_l$) to minimize $CV(h_l) = \frac{1}{N} \sum_{i=1}^{N} \left(Y_i^l - m^{-i}\left(X_{(i)}^m, h_l\right)\right)^2$, where $m^{-i}(x, h_l)$ is the estimate of the regression equation without using the observation $\left(Y_i^l, X_{(i)}^m\right)$ ($i = 1, 2, \cdots, N$). Therefore, we can consider the value of $h_l^{cv}$ ($l = 1, 2$) that is selected by the CV method as the best "predictive" estimate for $h_l$ ($l = 1, 2$). So, we propose to estimate the population variance when RSS and JPS are applied by using a concomitant variable by

$$\hat{\sigma}_N^2 = \frac{1}{kN} \sum_{i=1}^{kN} m\left(X_i, h_2^{cv}\right) - \left(\frac{1}{kN} \sum_{i=1}^{kN} m\left(X_i, h_1^{cv}\right)\right)^2.$$

**Remark 1**: It can be shown (see for example Takezawa, 2006, page 117) that $CV(h_l)$ can be computed efficiently by using the relation $CV(h_l) = \frac{1}{N} \sum_{i=1}^{N} \left(\frac{Y_i^l - m(X_i, h_l)}{1 - H_{ii}^{h_l}}\right)^2$, where the $H_{ii}^{h_l}$'s are the diagonal members of the hat matrix. In the simulation study in the next section, $h_l$ is selected from a sequence of values in the interval $[N^{-\frac{1}{4}}/3, 3N^{-\frac{1}{4}}]$ with steps of 0.01, where $N^{-\frac{1}{4}}$ is the bandwidth that is used by Frey (2011).

### 3. Monte Carlo Comparisons

In this section, we compare the proposed estimator with its leading competitors in RSS and JPS settings. For this purpose, we use the imperfect ranking model proposed by Dell and Clutter (1972), assuming $(Y, X)$ follows a standard bivariate normal distribution with correlation coefficient $\rho$. Then, we take $Y, \Phi(Y)$ and $-\ln(\Phi(Y))$ as the target variable, where $\Phi(.)$ is the distribution function of the standard normal distribution. So, we allow the relation between the target variable and the concomitant variable to be linear or non-linear and the parent distribution to be standard normal, uniform on (0,1) and exponential(1), respectively. Therefore we consider both symmetric and asymmetric distributions with bounded and unbounded supports. We compare the proposed estimator with its leading competitors in the literature in both balanced RSS and JPS settings, namely:

- Empirical variance estimator for balanced ranked set samples due to Stokes (1980), which has the form $\hat{\sigma}^2_{RSS} = \frac{1}{N-1}\sum_{i=1}^{N}(Y_i - \bar{Y})^2$.

- Variance estimator proposed by MacEachern et al. (2002) and Perron and Sinha (2004). It has the form $\hat{\sigma}^2_M = \frac{\sum_{r \neq s}\sum_i\sum_j (Y_{[r]i} - Y_{[s]j})^2}{2n^2k^2} + \frac{\sum_r\sum_i\sum_j (Y_{[r]i} - Y_{[r]j})^2}{2n(n-1)k^2}$ for balanced ranked set samples, where $Y_{[r]i}$ $(r=1,\cdots,k; i=1,\cdots n)$ is the $i$th observation with judgment rank $r$.

- Stratified variance estimator for judgment post stratified data, which has the form

$$\hat{\sigma}^2_{JPS} = \frac{1}{\sum_{r=1}^{k} I(n_r > 0)}\sum_{r=1}^{k}\bar{Y}^2_{[r]} I(n_r > 0) - \left(\frac{1}{\sum_{r=1}^{k} I(n_r > 0)}\sum_{r=1}^{k}\bar{Y}_{[r]} I(n_r > 0)\right)^2$$

; where $n_r$ is the number of observations with judgment rank $r$, $\bar{Y}_{[r]}$, and $\bar{Y}^2_{[r]}$ are the mean of the observations and the mean of the square of the observations with judgment rank $r$, respectively.

- Variance estimator proposed by Frey and Feeman (2013) for judgment post stratified data. Let $s_1 \geq \cdots \geq s_m > 0$ be the ordered sample sizes $n_1,\cdots,n_k$ (the number of observations with rank $r$), where $m \leq k$ is the number of non-empty post strata. With these reordered post strata, let $Y_{i1},\cdots,Y_{is_i}$ be the members of post-stratum $i$ with size $s_i$. Then Frey and Feeman (2013)'s variance estimator is $\hat{\sigma}^2_F = \sum_{i=1}^{m} w_{ii}\sum_{r<s}(Y_{ir} - Y_{is})^2 + \sum_{i<j} w_{ij}\sum_r\sum_s(Y_{ir} - Y_{js})^2$, where code for calculating the coefficients $w_{ij}$ $(i=1,\cdots,m; j=1,\cdots,s_i)$ is available on request from Frey and Feeman.

**Remark 2**: It is worth mentioning that these competing variance estimators are also applicable in case that RSS and JPS are done without measuring concomitant variable values.

**Remark 3**: An alternative sampling scheme when measuring a concomitant variable is easy and cheap is double sampling. In this sampling strategy, the researcher first draws $N_0$ sample units from the population and only measures the concomitant. Then, he draws a subsample of size $N$ from the first sample and measures the variable of interest in this subsample. Based on an anonymous referee's suggestion, we have compared the proposed variance estimator in the RSS design with its counterpart (denoted here by $\hat{\sigma}^2_{N,DS}$) for a double sampling design with $N_0 = nk^2$ and $N = nk$, which uses the same number of measurements of the concomitant variable and the variable of interest as the corresponding RSS design. It should be noted that $\hat{\sigma}^2_{N,DS}$ is exactly the same as $\hat{\sigma}^2_N$ in the JPS design.

In order to compare the different variance estimators in both RSS and JPS settings, we define the relative efficiency of each estimator $\tilde{\sigma}^2$ as:
$$RE = \frac{M\hat{S}E\left(\hat{\sigma}^2_{ref}\right)}{M\hat{S}E\left(\tilde{\sigma}^2\right)},$$

where $\hat{\sigma}^2_{ref}$ is the empirical variance estimator in the RSS design ($\hat{\sigma}^2_{RSS}$) and the stratified variance estimator ($\hat{\sigma}^2_{JPS}$) in the JPS design. We have generated 10000 random samples from the Dell and Clutter (1972) imperfect rankings model for $N = 15, 30, 45$; $k = 3, 5$ and $\rho = 0, 0.8, 1$ in both balanced RSS and JPS settings. The values of $\rho$ are selected to give random rankings ($\rho = 0$), perfect rankings ($\rho = 1$), and rankings which are good enough that both RSS and JPS sampling schemes are believed to give considerable improvement over SRS ($\rho = 0.8$). The results are presented in Tables 1 and 2.

**Table 1.** Estimated relative efficiencies of $\hat{\sigma}_M^2$, $\hat{\sigma}_N^2$ and $\hat{\sigma}_{N,DS}^2$ to $\hat{\sigma}_{RSS}^2$ for the balanced RSS scheme.

| Target Variable | | $Y$ | | | $\Phi(Y)$ | | | $-\ln(\Phi(Y))$ | | |
|---|---|---|---|---|---|---|---|---|---|---|
| | $\rho$ | $\hat{\sigma}_M^2$ | $\hat{\sigma}_N^2$ | $\hat{\sigma}_{N,DS}^2$ | $\hat{\sigma}_M^2$ | $\hat{\sigma}_N^2$ | $\hat{\sigma}_{N,DS}^2$ | $\hat{\sigma}_M^2$ | $\hat{\sigma}_N^2$ | $\hat{\sigma}_{N,DS}^2$ |
| | 0 | 0.98 | 1.05 | 1.05 | 0.99 | 0.95 | 0.90 | 0.99 | 1.05 | 1.03 |
| $(N,k)=(15,3)$ | 0.8 | 1.02 | 1.30 | 1.27 | 1.09 | 1.04 | 1.02 | 1.03 | 1.85 | 1.78 |
| | 1 | 1.00 | 1.92 | 1.90 | 1.14 | 2.03 | 1.77 | 1.05 | 2.72 | 2.63 |
| | 0 | 0.99 | 1.07 | 1.06 | 0.99 | 0.93 | 0.90 | 0.99 | 1.03 | 1.02 |
| $(N,k)=(15,5)$ | 0.8 | 1.09 | 1.34 | 1.22 | 1.11 | 1.03 | 1.00 | 1.05 | 1.87 | 1.84 |
| | 1 | 1.12 | 2.25 | 1.88 | 1.18 | 2.34 | 1.84 | 1.06 | 3.53 | 3.16 |
| | 0 | 0.99 | 1.01 | 1.00 | 0.99 | 0.92 | 0.93 | 0.99 | 1.01 | 1.01 |
| $(N,k)=(30,3)$ | 0.8 | 1.03 | 1.19 | 1.16 | 1.04 | 1.03 | 1.00 | 1.01 | 1.62 | 1.52 |
| | 1 | 1.04 | 2.02 | 1.98 | 1.07 | 2.35 | 2.23 | 1.02 | 2.59 | 2.50 |
| | 0 | 1.00 | 1.01 | 0.98 | 0.99 | 0.91 | 0.84 | 1.00 | 1.03 | 0.98 |
| $(N,k)=(30,5)$ | 0.8 | 1.04 | 1.18 | 1.12 | 1.05 | 1.01 | 0.90 | 1.02 | 1.79 | 1.70 |
| | 1 | 1.06 | 2.37 | 2.04 | 1.09 | 2.95 | 2.63 | 1.03 | 3.00 | 2.78 |
| | 0 | 0.99 | 0.99 | 0.98 | 0.99 | 0.93 | 0.92 | 0.99 | 1.01 | 1.00 |
| $(N,k)=(45,3)$ | 0.8 | 1.02 | 1.11 | 1.06 | 1.02 | 1.02 | 1.01 | 1.01 | 1.48 | 1.42 |
| | 1 | 1.03 | 2.11 | 2.10 | 1.04 | 2.43 | 2.40 | 1.01 | 2.56 | 2.46 |
| | 0 | 1.00 | 0.99 | 0.98 | 1.00 | 0.92 | 0.90 | 1.00 | 1.00 | 1.00 |
| $(N,k)=(45,5)$ | 0.8 | 1.02 | 1.09 | 1.04 | 1.03 | 1.01 | 0.99 | 1.01 | 1.54 | 1.45 |
| | 1 | 1.03 | 2.38 | 2.25 | 1.06 | 3.10 | 2.95 | 1.02 | 3.00 | 2.76 |

**Table 2.** Estimated relative efficiencies of $\hat{\sigma}_F^2$ and $\hat{\sigma}_N^2$ to $\hat{\sigma}_{JPS}^2$ for the JPS sampling scheme.

| Target Variable | | $Y$ | | $\Phi(Y)$ | | $-\ln(\Phi(Y))$ | |
|---|---|---|---|---|---|---|---|
| | $\rho$ | $\hat{\sigma}_F^2$ | $\hat{\sigma}_N^2$ | $\hat{\sigma}_F^2$ | $\hat{\sigma}_N^2$ | $\hat{\sigma}_F^2$ | $\hat{\sigma}_N^2$ |
| | 0 | 0.98 | 1.05 | 1.04 | 0.99 | 0.94 | 1.03 |
| $(N,k)=(15,3)$ | 0.80 | 1.02 | 1.30 | 1.06 | 1.06 | 0.99 | 1.74 |
| | 1 | 1.02 | 1.92 | 1.06 | 1.89 | 1.00 | 2.58 |
| | 0 | 1.03 | 1.14 | 1.11 | 1.04 | 0.97 | 1.06 |
| $(N,k)=(15,5)$ | 0.80 | 1.04 | 1.33 | 1.12 | 1.05 | 0.99 | 1.90 |
| | 1 | 1.04 | 2.11 | 1.14 | 2.17 | 0.99 | 3.42 |
| | 0 | 0.99 | 1.01 | 1.02 | 0.94 | 0.98 | 0.97 |
| $(N,k)=(30,3)$ | 0.80 | 1.01 | 1.16 | 1.02 | 1.00 | 0.98 | 1.60 |
| | 1 | 1.02 | 1.98 | 1.03 | 2.24 | 1.00 | 2.38 |
| | 0 | 1.05 | 1.09 | 1.08 | 1.02 | 1.01 | 1.06 |
| $(N,k)=(30,5)$ | 0.80 | 1.05 | 1.21 | 1.07 | 1.05 | 1.02 | 1.73 |
| | 1 | 1.06 | 2.21 | 1.04 | 2.84 | 1.03 | 2.89 |
| | 0 | 0.99 | 1.03 | 1.01 | 0.93 | 0.99 | 1.00 |
| $(N,k)=(45,3)$ | 0.80 | 1.00 | 1.12 | 1.01 | 1.01 | 1.00 | 1.38 |
| | 1 | 1.03 | 2.12 | 1.03 | 2.32 | 1.01 | 2.34 |
| | 0 | 1.04 | 1.06 | 1.05 | 0.99 | 1.02 | 1.02 |
| $(N,k)=(45,5)$ | 0.80 | 1.02 | 1.09 | 1.03 | 1.04 | 1.03 | 1.59 |
| | 1 | 1.02 | 2.44 | 1.02 | 3.06 | 1.00 | 2.80 |

Table 1 gives the results for balanced RSS. It is evident from this table that the proposed estimator performs well in comparison with its leading competitors and the performances of all estimators are almost the same when rankings are random ($\rho = 0$). When $Var(Y)$ is the parameter of interest, $\hat{\sigma}_N^2$ performs approximately twice as good as its competitors in the case of perfect rankings ($\rho = 1$). In the case of $\rho = 0.8$, $\hat{\sigma}_N^2$ is still the best, although its superiority decreases as sample size increases. The relative efficiencies of all estimators are almost the same when $\Phi(Y)$ is the target variable and $\rho \leq 0.8$. However, in the case of perfect rankings ($\rho = 1$), the proposed estimator is the best and its superiority increases as the set size ($k$) or the sample size ($N$) increases. In the case of $Var(-\ln(\Phi(Y)))$ being the parameter of interest, the proposed estimator is the best when rankings are good ($\rho \geq 0.8$). In this case, the relative efficiencies increase when $k$ increases and decrease when $N$ increases. It is also evident from Table 1 that the proposed estimator performs relatively better than its counterpart in double sampling when the rankings are good ($\rho \geq 0.8$).

The simulation results in the JPS setting are presented in Table 2. The pattern of this table is similar to that of Table 1. All estimators have similar performances when the rankings are random ($\rho = 0$) in most cases. When $Y$ is the target variable, $\hat{\sigma}_N^2$ is the best estimator when the rankings are good ($\rho \geq 0.8$), and in the case of perfect rankings ($\rho = 1$), it behaves approximately twice as good as the others and its relative efficiency increases as $N$ increases. The proposed estimator is the best in the case of perfect rankings ($\rho = 1$), when $Var(\Phi(Y))$ is the parameter of interest, and the improvement over other estimators increases as $N$ or $k$ increases. However, all estimators have almost the same performance when $\rho \leq 0.8$. In the case of $Var(-\ln(\Phi(Y)))$ being the parameter of interest, $\hat{\sigma}_N^2$ is the best one for $\rho \geq 0.8$. However, its relative efficiency decreases as sample size ($N$) increases or set size ($k$) decreases.

## 4. Conclusion

In this paper, we proposed a nonparametric variance estimator for ranked set samples and judgment post stratified data, when the ranking of the observations is done by measuring a concomitant variable. The proposed estimator was derived by using additional concomitant variable information and nonparametric kernel regression. The simulation results indicated that the new estimator typically performed substantially better than its competitors when the rankings are fairly good, while the performance was only slightly worse even under random rankings in almost all of the scenarios considered. As long as the correlation with the variable of interest can be assumed to be reasonably high, using the measured values of the concomitant variable in the proposed way can be expected to be highly advantageous compared to the competing estimators, which do not use the measured values, but only the ranking.

**Acknowledgements**

We thank Jesse Frey for providing us his R code for the calculation of the coefficients $w_{ij}$ that are used in the variance estimator proposed in Frey and Feeman (2013). We are also thankful to two anonymous referees for their valuable comments which improved an earlier version of this paper.